\documentclass[preprint,aps,nofootinbib,floatfix,superscriptaddress,longbibliography]{revtex4-1}
\usepackage{amsmath,amssymb,amsfonts,mathrsfs,bbm,braket,xcolor,graphicx,graphics,latexsym,placeins,epsfig,multirow,makecell}
\usepackage{array}
\newcolumntype{P}[1]{>{\centering\arraybackslash}p{#1}}
\usepackage{footmisc}
\usepackage{comment}
\usepackage{bbold}
\usepackage{enumitem}
\setdescription{leftmargin=8pt}
\usepackage{bbold}
\usepackage{hyperref}

\usepackage{graphicx,tabularx}
\usepackage{epsfig}
\usepackage{epstopdf}
\usepackage{amsfonts}
\usepackage{amssymb}
\usepackage{amsbsy}
\usepackage{amsmath}
\usepackage{mathrsfs}
\usepackage{latexsym}
\usepackage{bm}
\usepackage{color}
\usepackage{braket}
\usepackage{slashed}
\usepackage{pgfplots}
\usepackage{graphicx,subcaption}
\usepackage[normalem]{ulem}
\usepackage{mathrsfs}

\hypersetup{
	colorlinks=true,       
	linkcolor=blue,          
	citecolor=blue,        
	urlcolor=blue           
}
\DeclareMathOperator{\sech}{sech}
\usepackage{float}
\usepackage{amsmath,amssymb}

\newcommand{\ba}{\begin{align}}
\newcommand{\ea}{\end{align}}

\def\3nab{\tilde{\nabla}}

\def\be {\begin{equation}}
\def\ee {\end{equation}}
\def\ba {\begin{eqnarray}}
\def\ea {\end{eqnarray}}

\newcommand{\sfr}[2]
{{\textstyle\frac{#1}{#2}}}

\newcommand{\barray}{\begin{array}}
\newcommand{\earray}{\end{array}}

\newcommand{\bea}{\begin{eqnarray}}
\newcommand{\eea}{\end{eqnarray}}

\begin{document}

\title{Emergent Hawking Radiation and Quantum Sensing in a Quenched Chiral Spin Chain}
\author{Nitesh Jaiswal}
\email{niteshphy@iitb.ac.in}

\author{S. Shankaranarayanan}
\email{shanki@iitb.ac.in}

\affiliation{Department of Physics, Indian Institute of Technology Bombay, Mumbai 400076, India}

\begin{abstract}
We investigate the emergence and detection of Hawking radiation (HR) in a 1D chiral spin chain model, where the gravitational collapse is simulated by a sudden quantum quench that triggers a horizon-inducing phase transition. While our previous work~\cite{Shanky1} established that this model mimics black hole (BH) formation conditions even when the Hoop conjecture is seemingly violated, we here focus on the resulting stationary radiation spectrum and its detectability. By mapping the spin chain dynamics to a Dirac fermion in a curved $(1+1)$-dimensional spacetime, we analyze the radiation using two complementary approaches: field-theoretic modes and operational quantum sensors. First, using localized Gaussian wave packets to model realistic detectors, we find that the radiation spectrum exhibits deviations from the ideal Planckian form, while retaining robust Poissonian statistics that signal the loss of formation-scale information. 
Second, we introduce a qubit coupled globally to the chiral spin chain. Rather than acting as a localized point detector (which would break translational invariance), the qubit couples to the collective chiral excitations of the analogue spacetime, functioning as a global quantum sensor. We demonstrate that the decoherence dynamics of this macroscopic probe faithfully encode the thermal nature of the HR in the weak-coupling limit, while strong-coupling dynamics reveal non-Markovian memory effects driven by the horizon.
These results provide a clear operational protocol for distinguishing genuine analog HR from environmental noise in quantum simulation platforms.
\end{abstract}
\maketitle


\noindent \underline{\emph{Introduction:}} A central challenge in semiclassical and quantum gravity is the direct experimental verification of Hawking radiation (HR)~\cite{Parker:1975jm, Hawkingbh, Corley, Shankaranarayanan:2000qv,Shankaranarayanan:2000gb, Casadio:2001hp, FischerGB,RBMann, Kerner:2007rr, Kumar:2015bha, tapo2022,Suprit,Pal:2024qno}. While astrophysical observations, such as strong-field gravitational lensing, provide critical probes of the near-horizon geometry of black holes (BHs)~\cite{Mandal:2025tjx,Virbhadra:2022iiy}, the quantum dynamics of the horizon itself remain obscured by the weakness and delocalization of the radiation. Analogue gravity platforms have emerged as a powerful tool, as they enable controlled access to near-horizon quantum effects in laboratory settings ranging from fluids to optical systems~\cite{FischerBE, Barcelo:2005fc, Unruh:2012tz, Rajagopal:2014ewa, Steinhauer:2014dra, Steinhauer:2016hfa,  Banerjee:2019gli, Guo:2022goi}.

To extract these elusive signatures from the background of a many-body system, we turn to the paradigm of \emph{quantum sensing}~\cite{Degen:2016pxo,Pirandola:2018osg,DeMille:2024djy,Mercer:2025hfx,Zhang:2020skj}. Controlled spin-1/2 particles (qubits) have established themselves as highly sensitive probes of complex quantum environments, underpinning modern sensing technologies across molecular, atomic, and solid-state systems~\cite{Degen:2016pxo}. When coupled to an external bath, a qubit’s decoherence dynamics and population relaxation are governed by the spectral properties of the environment~\cite{Lewenstein}. This allows for the precise extraction of noise spectra~\cite{Clerk, Young} and Hamiltonian parameters~\cite{Degen:2016pxo} solely through measurements on the two-level probe.

Motivated by these capabilities, we employ a qubit coupled globally to the chiral chain, to probe the analogous radiation in the chiral spin chain~\cite{Cong:2021tnk}. In our previous work, we established that the chiral spin chain Hamiltonian can mimic gravitational collapse, satisfying black hole formation conditions even when the Hoop conjecture is violated~\cite{Shanky1}. In the present work, we focus on the implications for HR. Specifically, we investigate how the qubit's decoherence profile encodes the emergent thermal features of the horizon.

Our analysis proceeds in two steps. First, we examine the radiation spectrum directly using both idealized plane waves and physically realistic Gaussian wave packets. We show that while plane waves recover the standard thermal spectrum, localized packets reveal deviations akin to greybody factors~\cite{Birrell:1982ix,Wald:1995yp,Brout:1995rd, Unruh1, Don}, challenging the notion of HR as information-free. Second, we determine the operational conditions for detecting this temperature. We demonstrate that a qubit acts as a \emph{faithful thermometer} only in the weak-coupling regime, whereas strong coupling leads to thermalization with the bulk environment rather than the horizon. Finally, we discuss the statistical nature of the emission, showing that the radiation remains Poissonian regardless of the geometric ``Hoop" history, suggesting a universal erasure of formation scale information in the Hawking regime.

\begin{figure}[h]
    \centering
    \includegraphics[height=6.8cm, keepaspectratio]{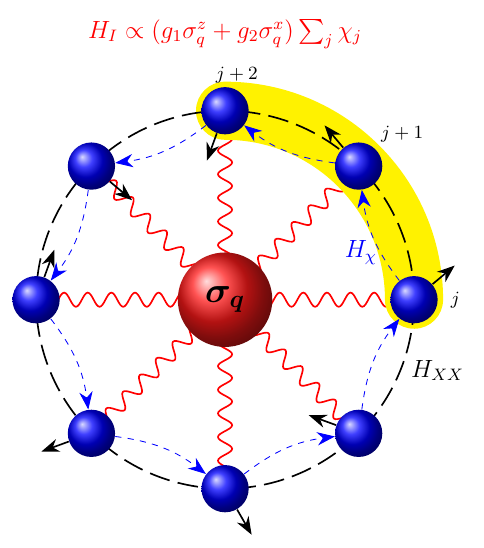}
    \caption{\label{fig:Model} Schematic diagram of the qubit–environment coupling for an $N=8$ chiral spin chain: the qubit (red) couples to the spins (blue) via the chirality operator $\chi_j$, while the spins interact through $XX$ and chiral terms. 
    }
\end{figure}

\noindent \underline{\emph{Model setup:}} To explore the dynamics of HR, we model the gravitational collapse as a sudden quantum quench \cite{Byju:2018eyb,Kaplanek:2020iay, bongs2023quantum,Covey,Balatsky}. We model the transition from a pre-horizon mass distribution to a BH using a time-dependent Hamiltonian $H(t)$. We assume the system initially evolves under a background Hamiltonian $H_0$ (representing the pre-collapse phase) and undergoes a sudden change at $t=t_0$ representing the formation of the horizon:
\begin{equation}
H(t) = H_0 + \Theta(t-t_0) H_h,
\end{equation}
where $\Theta(t-t_0)$ is the Heaviside step function and $H_h$ denotes horizon-inducing dynamics and the probe interaction. 
In our analogue model, the pre-horizon regime is governed by the standard $XX$ spin chain, given by~\cite{Shanky1}:
\begin{equation}\label{HXXeq}
H_0 \equiv H_{XX} = - \sum_{j=1}^{N} \frac{u(j,j+1)}{2} \left(\sigma_j^x\sigma_{j+1}^x+\sigma_j^y\sigma_{j+1}^y\right),
\end{equation}
where $\sigma_j^\alpha$ are Pauli operators at site $j$ satisfying periodic boundary conditions, and 
$u(j,j+1)\equiv u(|(j+1)-j|)$ is a site-dependent coupling constant depending only on the distance between nearest-neighbor sites. For $t \geq t_0$, the BH formation is captured by the activation of the chirality term $H_\chi$ and the coupling to an external probe $H_I$. The quench Hamiltonian is defined as $H_h = H_{\chi} + H_{I}$. The chirality term is:
\begin{equation}\label{Hchieq}
H_{\chi} = \sum_{j=1}^{N} (v/4) \, \chi_{j}~,\quad 
H_{I} = \sum_{j=1}^{N} (g_1\sigma_q^z + g_2\sigma_q^x)\chi_{j}/4,
\end{equation}
where $\chi_{j} = \vec{\sigma}_j \cdot (\vec{\sigma}_{j+1} \times \vec{\sigma}_{j+2})$ is three-spin chirality operator~\cite{Horner:2022sei, forbes2023exploring} and $v$ is a dimensionless coupling constant. The chirality operator induces handedness by breaking left–right symmetry and produces an imbalance in spin excitations, which cannot be generated by the \emph{inhomogeneous} $XX$ spin interactions alone. When activated by the quench, this term converts the inhomogeneity of the nearest-neighbor exchange coupling into an effective curvature in the emergent spacetime. To detect the thermal spectrum of HR, we couple the chain to a qubit via the interaction term $H_I$~\cite{Zanardi,Liu,Jaiswal_2022}. 
%
Here $\sigma_q^{x,z}$ act on the qubit and $g_{1,2}$ are coupling strengths. It is crucial to note the physical nature of the interaction Hamiltonian $H_I$. The qubit operators are summed over all lattice sites $N$, meaning the qubit interacts with the total macroscopic chirality of the chain, $\sum_j \chi_j$. Physically, this implies that the qubit does not act as a localized defect or a point-like Unruh-DeWitt detector, which would inevitably break the translational invariance of the lattice. Instead, the qubit is uniformly coupled to the bulk, acting as a global quantum sensor that probes the collective zero-momentum field excitations of the emergent curved spacetime. This global coupling preserves momentum conservation, allowing for an exact analytical diagonalization while capturing the macroscopic particle production generated by the horizon. As shown in Fig.~\eqref{fig:Model}, coupling allows the qubit to exchange energy with the chain, inducing decoherence that serves as a proxy for the thermal bath of HR.
%

\noindent \underline{\emph{Diagonalization and Dynamics:}} A key feature of our setup is the existence of a conserved quantity associated with the qubit. The operator $\mathcal{Q} \equiv g_1\sigma_q^z + g_2\sigma_q^x$ commutes with the total Hamiltonian, $[\mathcal{Q}, H(t)] = 0$. Consequently, the dynamics can be decomposed into independent sectors labeled by the eigenstates of $\mathcal{Q}$, denoted as $\ket{I_\pm}$:
\begin{equation}
\ket{I_\pm} = \frac{1}{\sqrt{1+|\mathscr{G}_\pm|^2}} 
\begin{bmatrix}
        \mathscr{G}_\pm\\
        1\\
    \end{bmatrix},\quad
\mathscr{G}_\pm = \frac{g_1 \pm \sqrt{g_1^2+g_2^2}}{g_2}.
\end{equation}
In each sector corresponding to eigenvalues $\pm\sqrt{g_1^2+g_2^2}$, the spin chain evolves under an effective Hamiltonian. Following the standard Jordan-Wigner and Fourier transformations~\cite{Shanky1,Jaiswal_2021}, the total Hamiltonian is diagonalized in momentum space (where $p\in [-\pi, \pi)$):
\begin{equation}
H(t) = \sum_{p} E_\pm(p) c^\dagger_p c_p,\end{equation}with the dispersion relation\begin{equation}E_\pm(p) = -2\,\mathcal{U}\cos(p) + V_{\cal E}^{\pm} \sin(2p)~,
\end{equation}
where $\mathcal{U}e^{ip}$ is the Fourier transform of $u(|(j+1)-j|)=\mathcal{U}$ (see Appendix~\ref{DiagAppendix} for details) and the effective chiral coupling is modified by the probe interaction: $V_{\cal E}^{\pm} = v \pm \sqrt{g_1^2+g_2^2}$.

\noindent \underline{\emph{Quantum Phase Transition and Time Evolution:}} Analysis of the ground-state energy density reveals a Quantum Phase Transition (QPT) driven by the competition between the nearest-neighbor hopping and the chiral interaction. A discontinuity in the second derivative of the energy density with respect to the effective coupling parameter $V_{\cal E}^{\pm}$ occurs at $|\mathcal{U}|=|V_{\cal E}^{\pm}|$, signaling a continuous QPT that separates the chiral and non-chiral phases.


 This critical threshold is central to our model, marking the boundary between distinct topological phases that we identify with the horizon formation condition. The physical origin of this QPT can be understood by drawing a parallel to Bosonic systems with higher spatial derivatives, such as the $z=3$ model studied in Ref.~\cite{Kumar:2016ucp}. There, it was shown that higher-order derivative terms --- corresponding to extended lattice neighbors --- can violate the area law of entanglement entropy and trigger a phase transition due to the accumulation of zero modes. Similarly, in our model, the chirality operator acts as a higher-order correction to the standard $XX$ chain dynamics. While $H_{XX}$ governs nearest-neighbor transport, $H_\chi$ induces three-site correlations that effectively modify the dispersion relation's topology. The critical point $|V_{\cal E}^{\pm}| = |\mathcal{U}|$ can be mapped to Lifshitz-like transition where the Fermi surface topology changes~\cite{revaz1999prb,fradkin2004ap,fradkinbook}, mimicking the causal disconnection of a horizon.

We assume the system is initially prepared in an uncorrelated product state at $t=t_0$:
\begin{equation}
\ket{\Psi(0)} = \ket{\psi_q(0)} \otimes \ket{\psi_E(0)},
\end{equation}
where $\ket{\psi_E(0)}$ is the vacuum of the pre-quench Hamiltonian $H_{XX}$, and the qubit is in a superposition $\ket{\psi_q(0)} = \sum_{\ell = -, +} c_{\ell} \ket{I_{\ell}}$. At $t=t_0$, a sudden quench is performed by abruptly changing the Hamiltonian $H_{XX}$ to the post-quench Hamiltonian $H(t)$. 
This leads to the generation of excitations and injected energy equivalent to a null shell collapse in the continuum limit~\cite{Zanardi,Jaiswal_2022}, triggering BH formation. Hence, for $t > t_0$, the evolution of the spin chain is driven by the conditional Hamiltonians 
$H_{\pm} = H_{XX} + H_{\chi} \pm \sum_{j=1}^{N}\sqrt{g_1^2+g_2^2}\, \chi_{j}/4 \ket{I_{\pm}}\bra{I_{\pm}}$, 
depending on the eigenstates of the qubit. Because the spin chain Hamiltonian can be decoupled in momentum space $p$, the environment state splits into two macroscopically distinct branches, 
$\ket{\psi_{E_{\pm}}(t)} = \prod_{p}e^{-i H_{\pm,p}\, t}\,\ket{\psi_E(0)}$,~\cite{Zanardi,Jaiswal_2022}
which factorize over the Fourier momentum modes $p$. The total wave function is a superposition of these conditionally evolved branches, making it a genuinely entangled state between the qubit and the spin chain:
\begin{equation}
|\Psi(t)\rangle = c_+ |I_+\rangle \otimes \left(\prod_p |\psi_{E_{+,p}}(t)\rangle\right) + c_- |I_{-}\rangle \otimes \left(\prod_p |\psi_{E_{-,p}}(t)\rangle\right),
\end{equation}
%
%
where $\ket{\psi_{E_{\pm, p}}(t)}$ represents the time-evolved state of the $p$-th momentum mode under the respective $H_{\pm,p}=E_\pm(p) c^\dagger_p c_p$ branch. 
%
%
This state encodes the \emph{thermal response} of the qubit due to background dynamics. In the rest of this work, we analyze this as an analogue of HR in BH.

\noindent \underline{\emph{Analogue BH Geometry:}} 
To understand the emergence of the analogue event horizon, we must map the spin chain dynamics to a geometric continuum limit. A sudden quantum quench injects energy into the system, as the pre-quench ground state is no longer an eigenstate of the post-quench Hamiltonian. This non-adiabatic transition inevitably generates a superposition of excited states, manifesting physically as a directed flow of quasi-particles (a \emph{flux of excitations}) that propagate ballistically outward~\cite{Calabrese:2016xau,Byju:2018eyb}. In the geometric language of analogue gravity, this sudden quench is mathematically equivalent to the gravitational collapse of a null shell of matter. The propagating quasi-particles correspond to the null energy flux that separates the spacetime into two distinct regions governed by different effective metrics, dynamically inducing an effective event horizon~\cite{Abajo-Arrastia:2010ajo}. The HR we derive arises from the propagation of vacuum fluctuations with respect to this new causal structure, independent of the background excitations flowing into the black hole.
As shown in Ref.~\cite{Shanky1}, the spin chain maps onto a Dirac fermion propagating in a $(1+1)-$dimensional geometry:
\begin{equation}i e_a^\mu \gamma^a (\partial_\mu + W_\mu) \psi(x) = 0, \label{Diraceq}
\end{equation}
where $\psi$ is rescaled by the tetrad determinant $\sqrt{|e|}$, $W_\mu$ is the spin connection, and $\gamma^a$ are the chiral gamma matrices (see Appendix~\ref{HRAppendix}). 

\begin{figure}
\centering
\includegraphics[width=0.4\textwidth]{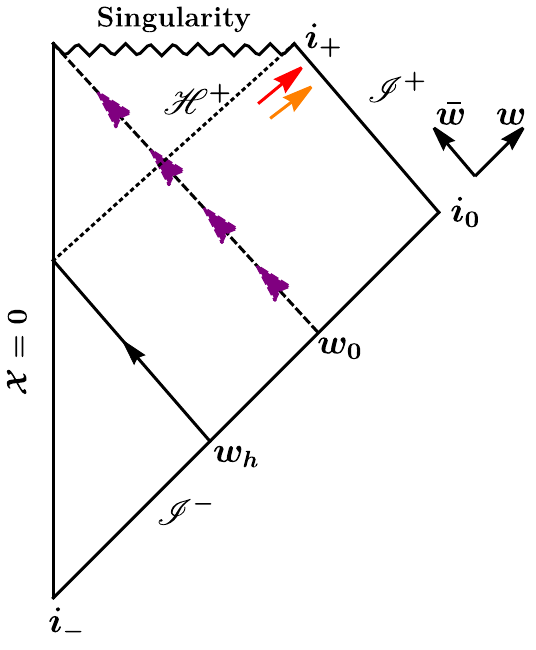}
        \caption{\label{fig:Penrose} Penrose diagram of the collapse process where a shock wave located at $w=w_0$ collapses to form a BH.}        
\end{figure}

The sudden quench protocol naturally partitions spacetime into two distinct causal regions separated by the collapse time $t_0$, as illustrated in the Penrose diagram (Fig.~\ref{fig:Penrose}). This mirrors the transition from an ``in" Minkowski region ($t < t_0$) to an ``out" BH region ($t \geq t_0$). The effective background metric is given by:
%
\begin{equation}
    ds^2 =\,\Biggl\{
  \begin{array}{@{}ll@{}}
    dt^2- d \mathcal{X}^2, & \,\,t< t_0 \\
    g(x) dt^2- 2 \sqrt{1 - g(x)}dt \,d\mathcal{X}  -d \mathcal{X}^2, &\,\, t\geq t_0
  \end{array}~,
\end{equation}
where $d\mathcal{X}=dx/\mathcal{U}(x)$ and $g(x) = 1 - {V_{\cal E}^{\pm}}^2/\mathcal{U}^2(x)$. Although, the nearest-neighbor exchange coupling $\mathcal{U}$ is an inhomogeneous quantity in both the pre and post-quench states, in the pre-quench regime this coupling merely rescales the local propagation velocity and can be absorbed into a coordinate redefinition, leaving the effective spacetime flat in $(t,\mathcal{X})$ coordinates. The quench, however, activates the chiral interaction, which explicitly breaks left–right symmetry and generates an imbalance in spin excitations, closely analogous to the charge separation underlying the chiral magnetic effect. This chiral term couples nontrivially to the existing inhomogeneities, and as a consequence the post-quench dynamics is no longer equivalent to that of flat space: the combined action of chirality and inhomogeneous $XX$ couplings curves the emergent spacetime and drives the system toward a topological QPT. The post-quench metric ($t \geq t_0$) corresponds to the Gullstrand-Painlev\'e coordinates of a spherically symmetric spacetime~\cite{Shankaranarayanan:2003ya}, with the spatial profile $\mathcal{U}$ acts as a position-dependent `speed of light' for the emergent Dirac fermions in the engineered BH geometry. 


\begin{table}[t]
\centering
\renewcommand{\arraystretch}{1.6} 
\begin{tabular}{ |c||c| } 
\hline
\textbf{Region} & \textbf{Null Coordinates}  \\
\hline

\multirow{2}{*}{\parbox[c]{7em}{\centering ``in''}} 
  & $w_{\rm in} = t + \mathcal{X}$,\quad $\bar{w}_{\rm in} = t - \mathcal{X}$, \\
& $\mathcal{X}=\sqrt{2}\,\sigma^{3/2}\sinh\!\left(\frac{x}{\sigma}\right)$. \\
\hline

\multirow{3}{*}{\parbox[c]{7em}{\centering ``out''}} 
  & $w_{\rm out} = \tau + \mathrm{X}$,\quad $\bar{w}_{\rm out} = \tau - \mathrm{X}$, \\

& $\tau=t-\int_{x_0}^x\frac{dy \,V_{e}^\pm}{\mathcal{U}^2(y) g(y)}$,\\

& $\mathrm{X} = \dfrac{i\sigma^2}{(V_{e}^\pm)^2\,\mathcal{X}_h}
\tan^{-1}\!\left(\dfrac{\mathcal{X}}{i\,\mathcal{X}_h}\right)$,\, $\mathcal{X}_h = \pm\sigma \sqrt{2(\sigma_c-\sigma)}$.\\[2mm]

\hline
\end{tabular}

\caption{Comparison of ``in'' and ``out'' null coordinates used in the analogue HR analysis.}
\label{tab:null_coords}
\end{table}

To analyze BH, we employ null coordinates to track the mixing of positive and negative frequency modes. The ``in" region is described by coordinates $(t, \mathcal{X})$, while the ``out" region utilizes $(\tau, \mathrm{X})$; the explicit transformations are detailed in Table~\ref{tab:null_coords}. For the spatial inhomogeneity, we adopt the profile~\cite{Shanky1}:
\begin{equation}\label{eq:formU}
\mathcal{U}(x) =  \text{sech}\left({x}/{\sigma}\right)/{\sqrt{2\sigma}},
\end{equation}
where $\sigma$ controls the width of the mass distribution. Horizon formation corresponds to $|V_{\cal E}^{\pm}| = |\mathcal{U}|$. For widths below a critical value $\sigma_c \equiv 1/(2{V_{\cal E}^{\pm}}^2)$, this condition yields two symmetric horizons at:
\begin{equation}x_{c\pm} = \pm \sigma \cosh^{-1}\left(\sqrt{\sigma_c/\sigma}\right).
\end{equation}
This double-horizon structure bears a striking resemblance to the Reissner-Nordstr\"om geometry, where the interplay of mass and charge similarly generates inner and outer horizons. Finally, to analyze the radiation, we employ two complementary probe states:
\begin{description}
\item[\textbf{Plane Waves}] These allow for analytic tractability and explicitly demonstrate the thermal nature of the spectrum (thermality).

\item[\textbf{Gaussian Wave Packets}] These localized states mimic realistic physical detectors, allowing us to track the propagation of Hawking quanta away from the horizon (detectability).
\end{description}

\noindent \underline{\emph{Analogue HR for Plane wave states:}} Fig.~\ref{fig:Penrose} summarizes the causal structure via the Penrose diagram. The shock wave transforms the flat ``in" region (past null infinity $\mathscr{I}^-$) into a BH geometry, forming an event horizon $\mathscr{H}^+$. Outgoing null rays originating near $\mathscr{H}^+$ experience exponential redshift before reaching future null infinity $\mathscr{I}^+$.

Positive-frequency ($f$) ingoing modes at $\mathscr{I}^-$ follow the standard form $\psi_f^{\rm in} \sim (e^{-i \bar{w}_{\rm in} f}, e^{-i w_{\rm in} f})^\text{T}$. Tracing the outgoing rays backward to the horizon reveals the characteristic logarithmic relation:
\begin{equation}
\bar{w}_{\rm out} \sim \frac{\sigma}{\sqrt{2(\sigma-\sigma_c)}}\log\left(\frac{w_{\rm in}-w_h}{w_{\rm in}-w_h-2\mathcal{X}_h}\right)~,
\end{equation}
where $\mathcal{X}_h=\pm\sigma \sqrt{2(\sigma_c-\sigma)}$. This logarithmic redshift mixes positive and negative frequency modes. The radiation spectrum is governed by the Bogoliubov coefficients $\alpha$ and $\beta$, which quantify the nontrivial mixing between positive and negative-frequency modes due to the time-dependent background (see Appendix~\ref{HRAppendix} for their rigorous definitions via the Dirac inner product). Computing the Bogoliubov coefficients 
yield the particle spectrum at late times:
\begin{equation}
\!\!\!\!\!
n_f^{P}=\int _0^\infty \!\!\!\! df^\prime |\beta_{ff^\prime}|^2 = \frac{1}{e^{{f}/{T_H}}+1}, \quad  T_H=\sqrt{\frac{\sigma_c-\sigma}{2\sigma^2 \sigma_c^2\pi^2}}~.
\label{Hawkingtemp}
\end{equation}
For $\sigma < \sigma_c$, the plane-wave analysis confirms a purely thermal Fermi-Dirac spectrum with temperature $T_H$.

\noindent \underline{\emph{Analogue HR for Gaussian states:}}  To model realistic detection, we employ localized Gaussian wave packets. We superpose the asymptotic plane-wave solutions with Gaussian envelopes ($\sim e^{-w^2}e^{-iwf}$), ensuring the states are spatially localized. The ratio of the resulting Bogoliubov coefficients is given by:
\begin{equation}
\frac{\alpha_{ff^\prime}^G}{\beta_{ff^\prime}^G}\sim e^{6iw_hf^\prime}\frac{\mathcal{D}_{-\nu}(z^\ast/\sqrt{2})}{\mathcal{D}_{-\nu}(z/\sqrt{2})}~,
\end{equation}
where $\mathcal{D}_{-\nu}$ is the Parabolic cylinder function, $z=-2 w_h-if^\prime$, and $\nu=1+if/(2\pi T_H)$. Unlike the delocalized plane-wave case, this ratio is generally non-thermal. 

%
\begin{figure}
     \centering
     \begin{subfigure}[b]{0.4\textwidth}
         \centering
         \includegraphics[width=\textwidth]{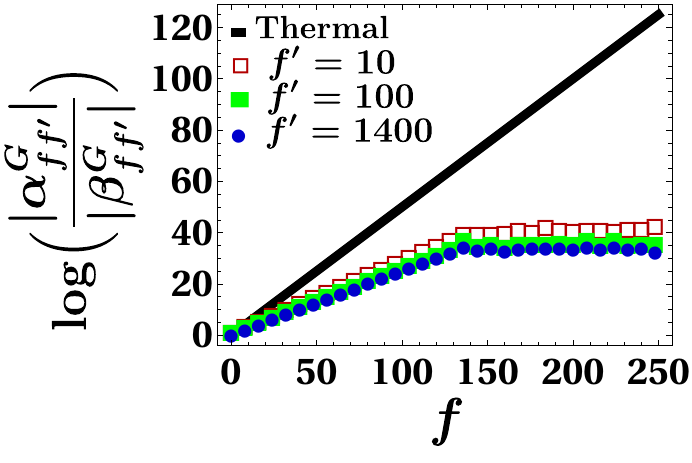}
     \end{subfigure}
  \hspace*{2cm}
     \begin{subfigure}[b]{0.36\textwidth}
         \centering
         \includegraphics[width=\textwidth]{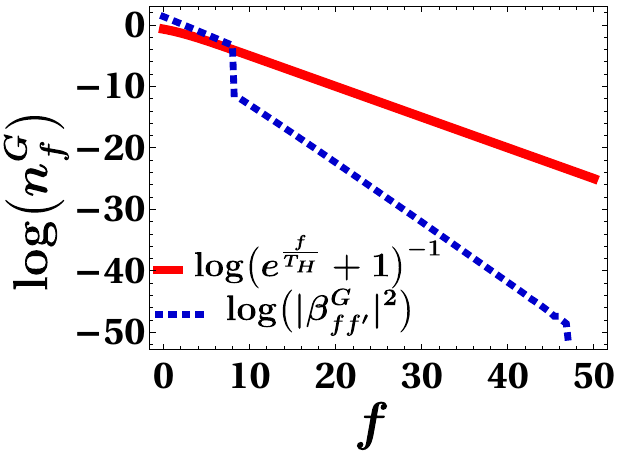}
     \end{subfigure}
     \caption{$\log\big(|\alpha_{ff^\prime}^G|/|\beta_{ff^\prime}^G|\big)$ as a function of outgoing frequency $f$ for various fixed incoming frequencies $f^\prime$ (left), and particle number spectrum $n_f^G$ as a function of $f$ comparing the thermal prediction with the localized Gaussian wave-packet probe (right).}
     \label{Bogoratio}
\end{figure}

Figure~\ref{Bogoratio} illustrates the effects of localization at fixed $T_H=2$ as determined from (\ref{Hawkingtemp}) with $\sigma=0.1014$ and $\sigma_c=0.1115$ is fixed by choosing $v=1$, $g_1=1$, and $g_2=1/2$. The left panel shows the logarithmic Bogoliubov ratio. At low frequencies, it follows the linear thermal scaling, but at high frequencies, it saturates to a constant. This indicates that physical detectors are insensitive to trans-Planckian modes. The right panel confirms that the particle number spectrum $n_f^G$ deviates from the ideal thermal prediction~\cite{ALKAC2025139783}. Thus, strict thermality is a feature of idealized infinite-duration measurements; realistic probes observe deviations at high energies. 
\begin{figure}
     \centering
     \begin{subfigure}[b]{0.4\textwidth}
         \centering
         \includegraphics[width=\textwidth]{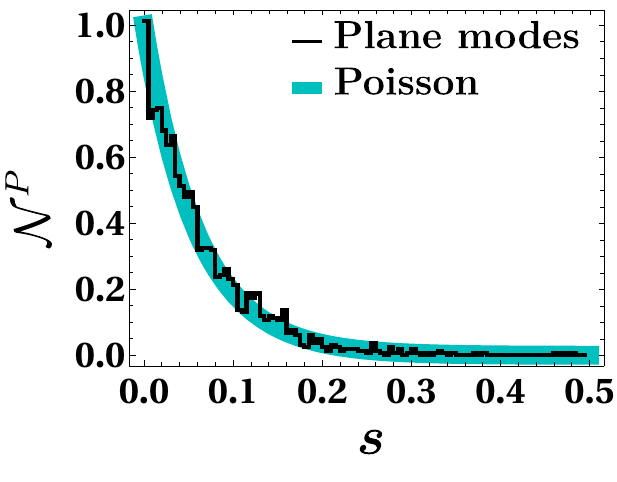}
     \end{subfigure}
  \hspace*{2cm}
     \begin{subfigure}[b]{0.4\textwidth}
         \centering
         \includegraphics[width=\textwidth]{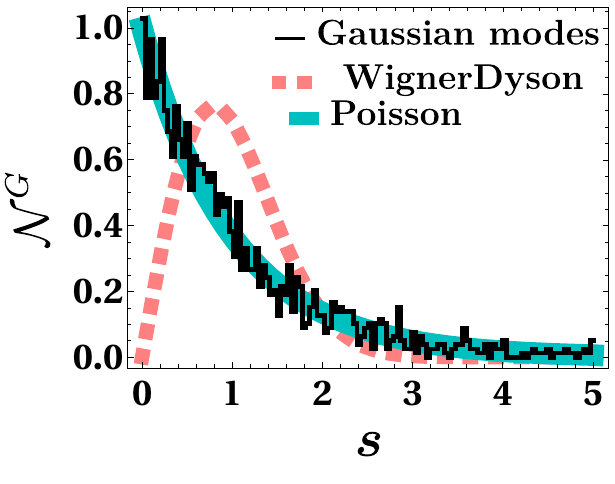}
     \end{subfigure}
     \caption{Poissonian statistics of Hawking emission for the plane-wave (left) and Gaussian wave-packets (right).}
     \label{Statisticsfig}
\end{figure}

Finally, we examine the statistical nature of the radiation. To systematically analyze the statistical properties of the emitted radiation and construct the distributions shown in Fig.~\ref{Statisticsfig}, we examine the level spacing statistics of the frequency modes. We first define a normalized probability density function for both the Plane ($P$) and Gaussian ($G$) wavepackets as:
\begin{equation}
\mathcal{P}_{P,G}(f)= {n_f^{P,G}} \Biggm/ {\int_0^\infty df \, n_f^{P,G}}~,\label{eq:normalized_prob}
\end{equation}
where $n_f^{P,G}$ represents the respective frequency spectrum (defined in Eq. \eqref{Hawkingtemp} for the plane-wave case). Using this probability density, a sequence of frequency samples is numerically generated and sorted in ascending order. From these ordered samples, we compute the consecutive frequency spacing variable, $s \equiv \langle n_f^{P,G} \rangle$. The quantity $\mathcal{N}^{P,G}$ plotted on the vertical axis of Fig. \eqref{Statisticsfig} is the numerically estimated probability density of this spacing variable $s$. It is constructed by histogramming the computed $s$ values and normalizing the area of the histogram to unity. Consequently, $\mathcal{N}^{P,G}$ does not possess a predefined analytic expression but is purely derived from the simulated spectral data.

To interpret these numerical results, we compare $\mathcal{N}^{P,G}$ against two standard analytical references. As clearly seen in Fig. \eqref{Statisticsfig}, the numerical data rigorously follows the analytic Poissonian distribution, $P(s) = e^{-s/s_0}$ (solid green line), which is the expected behavior for uncorrelated thermal emission. It distinctly deviates from the Wigner-Dyson distribution~\cite{Mehta-Book}, $P_{WD}(s) = (\pi s/2)e^{-\pi s^2/4}$ (dashed magenta line), which would otherwise indicate correlated level repulsion.

%
%
Contrary to Wigner–Dyson statistics, both plane-wave and Gaussian distributions are well-described by the Poisson form $e^{-s/s_0}$. This demonstrates that the statistical character of HR is robust: even when localization effects modify the spectrum, the emission process remains fundamentally Poissonian.

This is the key result of this work regarding which we want to discuss the underlying physical reason for the universal Poissonian behavior. HR originates from vacuum pair-production events which, in the absence of significant backreaction, are statistically independent and uncorrelated in time. This independence renders the emission a Poisson process, analogous to random tunneling events~\cite{Shankaranarayanan:2000qv,Shankaranarayanan:2003ya}. Since Gaussian wave packets are formed by linear superpositions of plane waves without introducing phase coherence, they preserve this intrinsic stochasticity. Thus, the Poissonian statistics reflect the fundamental randomness of vacuum fluctuations, which is robust against the spatial localization of the probe.

From the perspective of the microscopic spin chain, the distinction between these modes is physically interesting. Plane waves correspond to highly idealized, globally delocalized momentum eigenstates of the lattice. In contrast, Gaussian wavepackets represent physically realizable, localized spin excitations. While the kinematic existence of the horizon is fixed by the effective metric, our results demonstrate that the exact thermality of HR is an artifact of idealized plane waves. Localized wavepackets which represent the actual excitations a finite detector would absorb--inherently retain non-thermal correlations. 

It is important to note that the fundamental scattering dynamics of plane waves and the resulting greybody factors are well-established results in the standard literature of quantum field theory in curved spacetime~\cite{Birrell:1982ix, Wald:1995yp, Brout:1995rd, Unruh1, Don}. Our motivation for explicitly comparing plane-wave and Gaussian-wavepacket descriptions is not to present novel greybody physics, but rather to address the specific operational requirements of an analog quantum simulator. In this framework, plane-wave solutions are utilized to maintain analytic control and to unambiguously extract the underlying thermal Hawking spectrum. In contrast, the Gaussian-wavepacket formalism is necessary to faithfully model the finite-time, spatially localized excitation and detection protocols that are inherent to a realistic spin-chain experiment. This dual approach bridges the idealized analytical thermality of the BH horizon with the practical realities of localized measurements in condensed matter systems.

\begin{figure}
     \centering
     \begin{subfigure}[b]{0.4\textwidth}
         \centering
         \includegraphics[width=\textwidth]{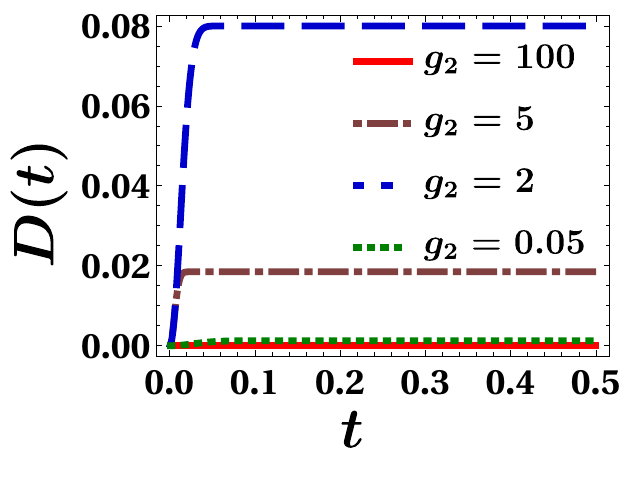}
     \end{subfigure}
  \hspace*{2cm}
     \begin{subfigure}[b]{0.38\textwidth}
         \centering
         \includegraphics[width=\textwidth]{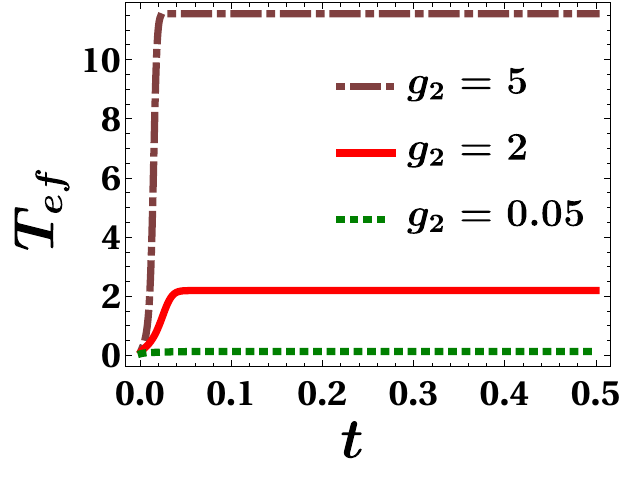}
     \end{subfigure}
     \caption{The dynamics of the decoherence factor (left) and the effective temperature (right) of the qubit for different values of $g_2$.}
     \label{Decoherencefig}
\end{figure}

\begin{figure}
     \centering     
         \includegraphics[width=0.55\textwidth]{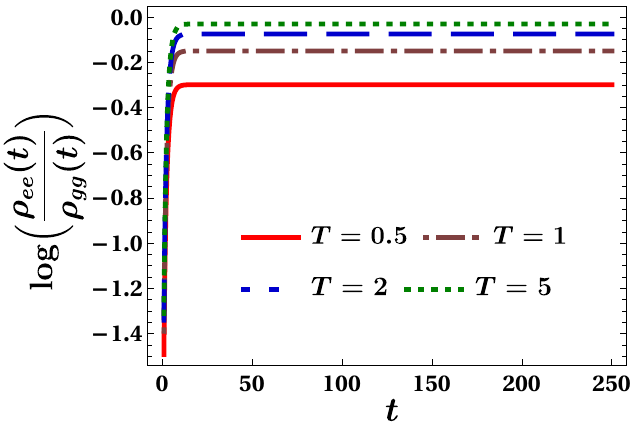}
        \caption{Time evolution of the logarithmic qubit population ratio for different values of the environmental temperature $T$ showing saturation at long times.}
     \label{spectraPopulationfig}
\end{figure}

\noindent \underline{\emph{Qubit as a quantum probe:}} 
Because the qubit couples to the global modes of the spin chain, its response reflects the macroscopic entanglement and particle production generated by the dynamical quench. In the weak-coupling (Markovian) limit, the qubit acts as a collective thermometer, relaxing to a steady state that faithfully registers the Hawking temperature of the bulk radiation. In contrast, the strong-coupling regime is dominated by measurement back-action and non-Markovian revivals, masking the purely thermal signature but revealing the robust entanglement structure of the post-quench vacuum.
\begin{description}
\item[\textbf{Strong coupling regime ($|g_2| > |\mathcal{U}|, |v|$)}] We first consider the qubit strongly coupled to the chain. By tracing out the environment, we compute the exact reduced density matrix $\rho_q(t)$ (see Appendix~\ref{QubitAppendix} for details). In this regime, the spin chain acts as a finite many-body reservoir. The interaction induces energy exchange, causing the qubit populations to relax to a steady state on a timescale $\tau_{0}\propto (g_1^2+g_2^2)^{-1/2}$ leading to rapid decoherence.
The decoherence function, defined as $D(t)=\prod_p|P_{ge}(p,t)|^2$, quantifies the magnitude of coherence present in the reduced density matrix $\rho_q(t)=\mathrm{Tr}_E \left[\,\ket{\Psi(t)}\bra{\Psi(t)}\,\right]$ obtained by tracing over the spin-chain modes $E$. The matrix elements in the basis $\{\ket{g}, \ket{e}\}$ are given by:
\begin{align}
    P_{gg}(p,t) &= \frac{2g_1^2+g_2^2\left( \cos \left(\Omega_Rt\right)+1\right)}{2(g_1^2+g_2^2)}~, \nonumber \\
    P_{eg}(p,t) &= \frac{g_1g_2(\cos \left(\Omega_Rt\right)-1)}{2(g_1^2+g_2^2)}-i\frac{g_2\sin \left(\Omega_Rt\right)}{2\sqrt{g_1^2+g_2^2}}~,
\end{align}
where $P_{ee}(p,t)=1- P_{gg}(p,t)$ and the offdiagonal coherence term is $P_{ge}(p,t)=P_{eg}^\ast(p,t)$. The Rabi frequency is defined as $\Omega_R=2 \sqrt{g_1^2+g_2^2} \sin{2p}$ and the total
population is the product over all quasi-momenta $p$.
As shown in Fig.~\ref{Decoherencefig}, {$D(t)$ decays sharply as the coupling between qubit and environment becomes stronger.} 
However, the effective temperature $T_{ef}$ extracted from the steady-state populations does not coincide with the Hawking temperature $T_H$. Instead, $T_{ef}$ depends non-trivially on the coupling parameters $g_{1,2}$ and the global properties of the chain. In this strong-coupling limit, the qubit effectively thermalizes with the entire spin-chain environment rather than selectively probing the horizon-induced modes. Consequently, a strongly coupled qubit is not a faithful probe of HR.

\item[\textbf{ Weak coupling regime ($|g_2| \ll |\mathcal{U}|, |v|$)}] Here, the spin chain acts as a Markovian thermal bath. The reduced qubit density matrix are governed by master rate equations determined solely by the bath spectral density $J(\omega)$ and the mode occupation~\cite{Breuer}:
\begin{equation}
\dot{\rho}_{ee} = - \Gamma_\downarrow \rho_{ee} + \Gamma_\uparrow \rho_{gg},
\end{equation}
where the rates $\Gamma_{\uparrow,\downarrow} \propto J(\omega)(e^{\pm \omega/T_H}+1)^{-1}$ follow the Fermi-Dirac distribution. These rates yield the reduced qubit dynamics for the excited-state population $\rho_{ee}(t)=f_D(\omega)\left(1-e^{-J(\omega)t}\right),$
and the ground-state population is simply given by $\rho_{gg}(t)=1-\rho_{ee}(t)$, showing relaxation toward a stationary thermal state. Unlike the strong coupling case, here the qubit is insensitive to microscopic details and probes only the thermal state of the Fermionic field. Solving the master equation confirms that the qubit relaxes to a stationary state $\rho_{ee}^s = (e^{\omega/T_H}+1)^{-1}$. Figure~\ref{spectraPopulationfig} illustrates the equilibration of the population ratio $\log(\rho_{ee}/\rho_{gg})$. The slope of this ratio in the steady state directly yields the Hawking temperature $T_H$. Thus, in the weak-coupling limit, the qubit functions as a faithful quantum thermometer, allowing for the experimental determination of the analogue Hawking temperature via simple population tomography.
\end{description}
%

%

\noindent \underline{\emph{Conclusions and Outlook:}} 
In this work, we have established a robust analogue of HR in a chiral spin chain modeled as a sudden quantum quench. Although the specific metric derived here arises from the spin chain couplings, the $\text{sech}(x)$ profile generates an effective geometry that shares the causal structure and multiple horizons of a Reissner-Nordstr\"om BH, making these insights applicable to astrophysical spacetimes. Furthermore, we observe that the HR is statistically insensitive to the `formation criteria'—the specific dynamical details of the quantum quench that created the horizon. While the spectral thermality fluctuates depending on whether the `in' and `out' vacua are probed via idealized plane waves or localized Gaussian wavepackets, the underlying emission remains robustly Poissonian. This memoryless statistical signature confirms that the core features of HR transcend both the microscopic UV completion and the specific dynamics of the gravitational collapse. 

Our work uniquely bridges abstract quantum field theory in curved spacetime with modern quantum information science. By introducing a qubit as a global quantum sensor, we provide a concrete, operational framework for detecting horizon-induced correlations via collective modes. This clarifies which aspects of HR are genuinely geometric and which are tied to measurement dynamics, offering a novel protocol for probing macroscopic relativistic entanglement without requiring localized defect modeling.

First, our analysis distinguishes between mathematical thermality and physical detectability. While idealized plane-wave modes recover a strictly thermal spectrum, we showed that physically realistic 
probes --- modeled as Gaussian wave packets --- exhibit deviations from the Planckian distribution. These deviations arise from the localization of the detector. 
This result challenges the conventional view of HR as a structureless, information-free thermal bath; instead, it suggests that measurement processes naturally imprint non-thermal features onto the observed spectrum, potentially encoding information about the detector-horizon interplay at finite scales.

Second, the universal nature of our results sheds light on the role of formation history. In our previous work, we showed that the geometric satisfaction of the Hoop conjecture is a necessary but insufficient condition for horizon formation in the percolation picture~\cite{Shanky1,bonnor1983hoop}. However, the present analysis reveals that the stationary HR is insensitive to these formation criteria. The emission statistics remain robustly Poissonian for both probes. This insensitivity is not a limitation of the model but a manifestation of the universal kinematics of the near-horizon region. The radiation effectively \emph{washes out} the memory of the formation scale (Hoop information), reinforcing the idea that HR acts as a late-time attractor that is independent of the microscopic details of the collapse.

Third, we emphasize that our model operates strictly in the continuum limit of the spin chain. Consequently, it reproduces the continuous Hawking spectrum characteristic of semiclassical gravity but remains blind to discrete lattice effects such as the Bekenstein-Mukhanov spectrum~\cite{Bekenstein:1995ju}. The quantization of horizon area, 
requires retaining the discrete lattice structure of the chain or use other probes like in fluid-gravity correspondence~\cite{Cropp:2016ajh}.

Finally, we have identified the operational conditions required to measure this effect in a quantum simulation. We showed that a qubit acts as a faithful thermometer only in the weak-coupling regime, where it essentially probes the spectral density of the bath. In contrast, strong coupling induces complex many-body entanglements where the qubit thermalizes with the entire environment rather than the horizon specifically. This distinction is critical for future experimental realizations of analogue gravity in quantum spin systems, providing a clear protocol for distinguishing genuine Hawking temperatures from spurious environmental heating, especially in setups that mimic cosmological expansion~\cite{Chandran:2025azu}.

While the present work establishes the operational utility of a global quantum sensor, a natural future direction is to investigate a strictly localized probe coupled to a single lattice site $j_0$. Because such a localized coupling breaks translational invariance, this would require analyzing the horizon dynamics through the lens of open quantum systems impurity physics—such as a Caldeira-Leggett model with a localized defect to explore how exact spatial sampling affects the detected Hawking spectrum.

\noindent {\bf Acknowledgments:} The authors thank S. Bera, S.~Bhattacharya,  I.~Chakraborty, K. Hari, P. George Christopher, T.~Parvez, and T.~Sarkar for discussions and comments on the earlier draft. We thank the anonymous referee for the constructive comments and suggestions, which helped improve the clarity and presentation of this manuscript. The work is supported by ANRF/ARG/2025/001514/PS.



\appendix

\section{The Model and its Diagonalization}
\label{DiagAppendix}

The total time-dependent Hamiltonian $H(t)$ is composed of three parts: the conventional $XX$ spin-chain Hamiltonian, a chirality operator term, and an interaction term describing the coupling between the qubit and the chiral sector:
\begin{equation}
    H(t)= - \sum_{j=1}^{N} \frac{u(j,j+1)}{2}\left(\sigma_j^x\sigma_{j+1}^x+\sigma_j^y\sigma_{j+1}^y\right)+ \Theta(t-t_0)\sum_{j=1}^{N} \left[\frac{v}{4} \, \chi_{j}+(g_1\sigma_q^z + g_2\sigma_q^x)\frac{\chi_{j}}{4}\right]~.
\end{equation}
Here, $u(j,j+1)\equiv u(|(j+1)-j|)$ is a site-dependent coupling constant depending only on distance between the nearest-neighbor sites, $\chi_{j} = \vec{\sigma}_j \cdot (\vec{\sigma}_{j+1} \times \vec{\sigma}_{j+2})$ denotes the three-spin chirality operator, $v$ is a dimensionless coupling constant, $\sigma_q^{x,z}$ are the Pauli matrices acting on the qubit, and $g_{1,2}$ represent the corresponding coupling strengths. At $t=t_0$, a sudden quench leads to changing the XX spin-chain Hamiltonian to the post-quench Hamiltonian $H(t)$. Since the quench occurs on a time scale much shorter than the intrinsic dynamical scales of the system, the initial ground state does not match the ground state of the new Hamiltonian, leading to the generation of excitations and injected energy~\cite{Zanardi,Jaiswal_2022}. This sudden quench is equivalent to a null shell collapse in the continuum limit of the above spin-chain model, triggering black hole formation in the emergent spacetime. 

We first carry out the diagonalization of the environment Hamiltonian $H_{E}$ comprising of the standard $XX$ spin chain and the three-spin chirality term using Ref.~\cite{Horner:2022sei,Shanky1}. We employ a Jordan–Wigner transformation, which provides an exact mapping between the spin-1/2 Pauli operators and a set of spinless fermionic operators defined as
\begin{equation}
    \sigma_{j}^{+}=\exp{\left(-i\pi\sum_{l=1}^{j-1}c_{l}^{\dagger}c_{l}\right)}c_{j}^{\dagger}~,\quad \sigma_{j}^{-}=\exp{\left(i\pi\sum_{l=1}^{j-1}c_{l}^{\dagger}c_{l}\right)}c_{j}~,\quad  \sigma_{j}^{z}=1-2c_{j}^{\dagger}c_{j}~,
\end{equation}
where $\sigma_{j}^{\pm}= (\sigma_j^x\pm \sigma_j^y)/2$, while $c_j$ and $c^\dagger_j$ are fermionic annihilation and creation operators, respectively satisfying anti-commutation relations $\{c_j,c_k^\dagger\}=\delta_{jk}$ and $\{c_j,c_k\}=\{c_j^\dagger,c_k^\dagger\}=0$. Using this we can write the environment Hamiltonian
\begin{equation}
    H_{E}=\sum_{j=1}^{N}\left[-u(|(j+1)-j|)c^\dagger_jc_{j+1}+\frac{iv}{2}\left(c^\dagger_jc_{j+1}\sigma_{j+2}^z+c^\dagger_{j+1}c_{j+2}\sigma_{j}^z-c^\dagger_jc_{j+2}\right)\right]+H.C.~.
\end{equation}
The above Hamiltonian contains four-fermion interaction terms rendering it interacting and not directly diagonalizable. To proceed, we employ a mean-field approximation in which the operators $\sigma_{j}^z$ are replaced by their expectation value $Z$, assumed uniform due to translational invariance. Imposing self-consistency by requiring $\braket{\Omega(Z)|\sigma_{j}^z|\Omega(Z)}=Z$, where $\ket{\Omega(Z)}$ is the ground state of the resulting quadratic Hamiltonian, we obtain the solution $Z=0$~\cite{Horner:2022sei} using particle-hole symmetry and Jordan-Wigner transformation. After that, the environment Hamiltonian can be exactly diagonalized using the discrete Fourier transformation
\begin{equation}
    c_j = \frac{1}{\sqrt{N}}\sum_{p\in [-\pi,\pi)} e^{ipj}c_p~,\quad c_j^\dagger = \frac{1}{\sqrt{N}}\sum_{p\in [-\pi,\pi)} e^{-ipj}c_p^\dagger~,
\end{equation}
where the quasi-momentum $p=\frac{2\pi\lambda}{N}$ for $\lambda\in \mathbb{Z}$ yielding
\begin{equation}
    H_{E} = -\frac{1}{N} \sum_{p,\,p^\prime} \sum_{j=1}^{N}u(|(j+1)-j|)e^{i(p^\prime (j+1)-pj)}c_p^\dagger c_{p^{\prime}} +\sum_{p} \frac{iv}{2} e^{2ip}c_p^\dagger c_p+ H.C.~.
\end{equation}
Rewriting the above Hamiltonian in the diagonalized form:
\begin{equation}
     H_{E}=\!\! \sum_{p\in [-\pi, \pi)} \!\! E(p) c^\dagger_p c_p~,\quad E(p)= -2\,\mathcal{U}\cos(p) + v \sin(2p)~,
\end{equation}
where $\mathcal{U}=u(|(j+1)-j|)$ and $E(p)$ is the dispersion relation of the environment Hamiltonian.

To diagonalize the total Hamiltonian $H(t)$ by incorporating the contribution from the qubit sector, we first observe that the operator $\mathcal{Q} \equiv g_1\sigma_q^z + g_2\sigma_q^x$ commutes with the total Hamiltonian, $[\mathcal{Q}, H(t)] = 0$. As a result, $\mathcal{Q}$ is a conserved quantity and can be treated as a constant within each sector. Its eigenvalues are $\pm \sqrt{g_1^2+g_2^2}$, with corresponding eigenstates given by~\cite{Zanardi,Liu}:
\begin{equation}
\ket{I_\pm} = \frac{1}{\sqrt{1+|\mathscr{G}_\pm|^2}} 
\begin{bmatrix}
        \mathscr{G}_\pm\\
        1\\
    \end{bmatrix}, \quad 
\mathscr{G}_\pm = \frac{g_1 \pm \sqrt{g_1^2+g_2^2}}{g_2}~.
\end{equation}
This allows us to replace the qubit subsystem by its spectral decomposition:
\begin{equation}
    g_1\sigma_q^z + g_2\sigma_q^x \equiv \sum_{\ell = \pm} \ell \sqrt{g_1^2+g_2^2}\ket{I_{\ell}}\bra{I_{\ell}}~,
    \label{qubiteigenvalue}
\end{equation}
where $\ket{I_{\ell}}$ denote the corresponding qubit eigenstates. As a consequence, the full Hamiltonian decomposes into two independent sectors labeled by $\ell = \pm$. 
Hence, the total Hamiltonian can be written in diagonal form as
\begin{equation}
H(t) = \sum_{p} E_\pm(p) c^\dagger_p c_p~,\quad E_\pm(p) = -2\,\mathcal{U}\cos(p) + V_{\cal E}^{\pm} \sin(2p)~,
\end{equation}
where $V_{\cal E}^{\pm} = v \pm \sqrt{g_1^2+g_2^2}$ and $E_\pm(p)$ is the dispersion relation for the total Hamiltonian. 

The ground state energy density of the total Hamiltonian in the thermodynamic limit is:
\begin{equation}
    \rho_{\pm} = \frac{1}{2\pi}\int_{p:E_\pm(p)<0} dp E_\pm(p)\quad = \,\Biggl\{
  \begin{array}{@{}ll@{}}
    \frac{2\mathcal{U}}{\pi}, & V_{\cal E}^{\pm}\leq \mathcal{U} \\
    -\frac{\mathcal{U}^2 + (V_{\cal E}^{\pm})^2}{V_{\cal E}^{\pm} \pi}, & V_{\cal E}^{\pm}>\mathcal{U}
  \end{array}~.
\end{equation}
A discontinuity in the second-order derivative of $\rho_{\pm}$ with respect to the interaction strength parameter $V_{\cal E}^{\pm}$ signals a second-order QPT at $|V_{\cal E}^{\pm}|=|\mathcal{U}|$~\cite{Shanky1}. For $|V_{\cal E}^{\pm}|>|\mathcal{U}|$, the system is in the chiral phase, while $|V_{\cal E}^{\pm}|<|\mathcal{U}|$ resembles the non-chiral phase.

\section{Analogue Hawking Radiation}
\label{HRAppendix}

\subsection{Plane Wave States}
The covariant Dirac equation for massless spinor fields in curved spacetime is given by~\cite{Shanky1}:
\begin{equation}
    i e_a^\mu \gamma^a (\partial_\mu + W_\mu) \psi(x) = 0, \label{Diraceq}
\end{equation}
where $W_\mu$ denotes the spin connection, and $\gamma^a$ are the Dirac gamma matrices. In the Dirac representation, these are chosen as
\begin{equation}
    \gamma^0=\begin{pmatrix}1&0\\
    0&-1\end{pmatrix}~,\quad \gamma^1=\begin{pmatrix}0&-i\\
    -i&0\end{pmatrix}~.
\end{equation}
The gamma matrices obey the anti-commutation relation $\{\gamma^{a}, \gamma^{b}\}=2\eta^{ab}$ with the Minkowski metric $\eta^{ab}=\text{diag}(1,-1)$. To solve Eq.~(\ref{Diraceq}), it is convenient to work in the chiral representation, in which the left and right-moving modes decouple. The gamma matrices in the chiral basis are obtained via the unitary transformation $S$: $\gamma^\mu_c=S^{-1}\gamma^\mu S$ yielding
\begin{equation}
   \gamma^0_c=\begin{pmatrix}0&1\\
    1&0\end{pmatrix}~,\quad \gamma^1_c=\begin{pmatrix}0&-1\\
    1&0\end{pmatrix}~,\quad \text{with} \quad S=\frac{1}{\sqrt{2}}\begin{pmatrix}1&1\\
    i&-i\end{pmatrix}~. 
\end{equation}
In the flat spacetime ``in'' region, it is natural to introduce coordinates $(t, \mathcal{X})$ with the spatial coordinate defined through $d\mathcal{X}=dx/\mathcal{U}(x)$. In these coordinates, the Dirac equation and its plane wave solution are
\begin{equation}
    i( \gamma^0_c \partial_t+ \gamma^1_c\partial_{\mathcal{X}})\psi_f^{\text{in}}(t,\mathcal{X})=0~,\quad \psi_f^{\text{in}}(t,\mathcal{X})\sim \begin{pmatrix}
        e^{-i\bar{w}_{\text{in}}f}\\
        e^{-iw_{\text{in}}f}
    \end{pmatrix}~,
    \label{InDirac}
\end{equation}
for positive frequency ingoing modes with the ingoing null coordinates $w_{\text{in}} = t + \mathcal{X}$, $\bar{w}_{\text{in}} = t - \mathcal{X}$. In contrast, the ``out'' region corresponds to a curved spacetime and is described by coordinates $( \tau , \mathrm{X})$. The Dirac equation and its plane wave solution in the ``out'' region are
\begin{equation}
    i\left( \gamma^\tau_c \partial_\tau+ \gamma^{\mathrm{X}}_c\partial_{\mathrm{X}}+\gamma^\tau_c W_\tau+\gamma^{\mathrm{X}}_c W_\mathrm{X}\right)\psi_f^{\text{out}}(\tau,\mathrm{X})=0~,\quad \psi_f^{\text{out}}(\tau,\mathrm{X})\sim \begin{pmatrix}
        e^{-i\bar{w}_{\text{out}}f}\\
        e^{-iw_{\text{out}}f}
    \end{pmatrix}~,
    \label{OutDirac}
\end{equation}
for positive frequency outgoing modes with the outgoing null coordinates $w_{\text{out}} = \tau + \mathrm{X}$, $\bar{w}_{\text{out}} = \tau - \mathrm{X}$. The explicit coordinate transformations take the forms
\begin{equation}
\tau=t-\int_{x_0}^x\frac{dy \,V_{\cal E}^{\pm}}{\mathcal{U}^2(y) g(y)}~,\quad \mathrm{X} = \frac{i\sigma^2}{(V_{\cal E}^{\pm})^2\,\mathcal{X}_h}
\tan^{-1}\!\left(\frac{\mathcal{X}}{i\,\mathcal{X}_h}\right)~, \quad
\mathcal{X}_h = \pm\sigma \sqrt{2(\sigma_c-\sigma)}~.
\end{equation}
In these coordinates, the curved-spacetime gamma matrices in the chiral representation are
\begin{equation}
    \gamma_c^\tau =\frac{1}{g(\mathrm{X})}\begin{pmatrix}
        0&& 1+\frac{V_{\cal E}^{\pm}}{\mathcal{U}(\mathrm{X})}\\
        1-\frac{V_{\cal E}^{\pm}}{\mathcal{U}(\mathrm{X})}&&0
    \end{pmatrix}~,\quad 
    \gamma_c^{\mathrm{X}}=\frac{1}{g(\mathrm{X})}\begin{pmatrix}
        0&& -1-\frac{V_{\cal E}^{\pm}}{\mathcal{U}(\mathrm{X})}\\
        1-\frac{V_{\cal E}^{\pm}}{\mathcal{U}(\mathrm{X})}&&0
    \end{pmatrix}~,
\end{equation}
and the corresponding spin-connection components entering the Dirac equation are
\begin{align}
    W_\tau &=\frac{(V_{\cal E}^{\pm})^2 \,\mathcal{U}^\prime(\mathrm{X})}{2\,\mathcal{U}^3(\mathrm{X})g(\mathrm{X})}\gamma^0~,\quad W_{\mathrm{X}}=\frac{V_{\cal E}^{\pm} \,\mathcal{U}^\prime(\mathrm{X})}{2\,\mathcal{U}^2(\mathrm{X})g(\mathrm{X})}\gamma^0~,\notag\\
    \mathcal{U}(\mathrm{X})&=\frac{1}{\sqrt{2 \sigma}} \left[1+\left(1-\frac{1}{2 \sigma {V_{\cal E}^{\pm}}^2}\right) \cot^2\left(\frac{\mathrm{X} V_{\cal E}^{\pm} \sqrt{2 \sigma {V_{\cal E}^{\pm}}^2-1}}{\sigma }\right)\right]^{-1/2}.
\end{align}
The calculation of Hawking radiation crucially relies on tracing outgoing field modes backward in time toward the horizon. The above null coordinates play a central role in tracing outgoing modes backward toward the horizon and in evaluating the Bogoliubov coefficients responsible for Hawking radiation. As emphasized in Ford’s review~\cite{Ford:1997hb}, this backward propagation reveals the exponential redshift experienced by outgoing rays near the horizon:
\begin{equation}
\bar{w}_{\text{out}} \sim \frac{\sigma}{\sqrt{2(\sigma-\sigma_c)}}\log\left(\frac{w_{\text{in}}-w_h}{w_{\text{in}}-w_h-2\mathcal{X}_h}\right)~,
\end{equation}
which in turn governs the nontrivial Bogoliubov mixing between positive and negative-frequency modes. In the context of Hawking radiation, Bogoliubov coefficients relate the vacuum defined in the distant past (``in'' region) to that seen by observers in the far future (``out'' region). They encode how gravitational collapse and horizon formation mix positive and negative-frequency modes, causing the initial vacuum to appear as a particle-filled state and leading to particle creation. These coefficients $\beta_{ff^\prime}$ and $\alpha_{ff^\prime}$ are defined by the Dirac inner product between the ``in'' and ``out'' modes
\begin{eqnarray}
   \beta_{ff^\prime} &\equiv & (\psi_f^{\text{out}},\,\psi_{f^\prime}^{\text{in}})_D=\int_\Sigma d\Sigma_\mu \bar{\psi}_f^{\text{out}}\gamma^\mu \psi_{f^\prime}^{\text{in}}~,\notag\\
   \alpha_{ff^\prime}
   &\equiv&
   (\psi_f^{\text{out}},\,\psi_{f^\prime}^{{\text{in}}^\dagger})_D
   =\int_\Sigma d\Sigma_\mu \bar{\psi}_f^{\text{out}}\gamma^\mu \psi_{f^\prime}^{{\text{in}}^\dagger}~,
\end{eqnarray}
where $d\Sigma_\mu=n_\mu d\Sigma$ is the directed surface element, and $n_\mu$ is the unit normal to the Cauchy surface $\Sigma$. The above integral is dominated by contributions from the vicinity of the horizon,
\begin{equation}
    \beta_{ff^\prime}\sim \int_{-\infty}^{w_h} dw_{\text{in}}\,\, e^{i w_{\text{out}}f}e^{-i w_{\text{in}}f^\prime}=-e^{-i w_{h}f^\prime}(if^\prime)^{-1-\frac{i f}{2\pi T_H}}\,\Gamma\left(1+\frac{i f}{2\pi T_H}\right)~,
\end{equation}
with $T_H=\sqrt{\frac{\sigma_c-\sigma}{2\sigma^2 \sigma_c^2\pi^2}}$ is the Hawking temperature and $n_\mu=(1,0)$ for a surface with constant $\bar{w}$. Similarly one can also obtain the other Bogoliubov coefficient, $|\alpha_{ff^\prime}|=e^{f/2 T_H}|\beta_{ff^\prime}|$. Then, using the normalization condition $\int_0^\infty df^\prime\left(|\alpha_{ff^\prime}|^2+|\beta_{ff^\prime}|^2\right)=1$, we obtain the particle spectrum at late times as
\begin{equation}
   n_f^P= \int _0^\infty \!\!\!\! df^\prime |\beta_{ff^\prime}|^2 = \frac{1}{e^{{f}/{T_H}}+1}~.
\end{equation}

\subsection{Gaussian Wave Packets}

Having established the plane-wave solutions of the Dirac equation in the ``in'' and ``out'' regions, we now construct localized Gaussian wave packet solutions. These are defined for the ``in'' and ``out'' regions respectively as
\begin{equation}
    \psi_{f,G}^{\text{in}} \sim 
    \begin{pmatrix}
    e^{-\bar{w}_{\text{in}}^2}e^{-i \bar{w}_{\text{in}} f}\\
    e^{-w_{\text{in}}^2}e^{-i w_{\text{in}} f}
    \end{pmatrix},\quad\quad \psi_{f,G}^{\text{out}} \sim 
    \begin{pmatrix}
    e^{-\bar{w}_{\text{out}}^2}e^{-i \bar{w}_{\text{out}} f}\\
    e^{-w_{\text{out}}^2}e^{-i w_{\text{out}} f}
    \end{pmatrix},
\end{equation}
for positive frequency modes, where the subscript $G$ stands for Gaussian. Since these Gaussian wave packets are constructed in analogy with the plane-wave modes, the Bogoliubov transformation relating ingoing and outgoing solutions proceeds in complete analogy with the plane-wave case. Consequently, the Bogoliubov coefficients are
\begin{align}
    \beta_{ff^\prime}^G &\sim \int_{-\infty}^{w_h} dw_{\text{in}}\,\, e^{i w_{\text{out}}f}e^{-i w_{\text{in}}f^\prime} e^{-w_{\text{out}}^2-w_{\text{in}}^2} \nonumber \\
    &= (-1)^{\nu +1}e^{-i w_{h}f^\prime}e^{-w_h^2}\,\Gamma(\nu)2^{-\frac{\nu}{2}}e^{\frac{z^2}{8}}\mathcal{D}_{-\nu}\left(\frac{z}{\sqrt{2}}\right)~,
\end{align}
and 
\begin{align}
    \alpha_{ff^\prime}^G &\sim \int_{-\infty}^{w_h} dw_{\text{in}}\,\, e^{i w_{\text{out}}f}e^{i w_{\text{in}}f^\prime} e^{-w_{\text{out}}^2-w_{\text{in}}^2} \nonumber \\
    &= (-1)^{\nu +1}e^{i w_{h}f^\prime}e^{-w_h^2}\,\Gamma(\nu)2^{-\frac{\nu}{2}}e^{\frac{{z^\ast}^2}{8}}\mathcal{D}_{-\nu}\left(\frac{z^\ast}{\sqrt{2}}\right)~,
\end{align}
where $\mathcal{D}_{-\nu}$ is the Parabolic cylinder function, $z=-2 w_h-if^\prime$, and $\nu=1+if/(2\pi T_H)$. Then, the ratio of Bogoliubov coefficients for Gaussian wave packets can be evaluated as
\begin{equation}
\frac{\alpha_{ff^\prime}^G}{\beta_{ff^\prime}^G}\sim e^{6iw_hf^\prime}\frac{\mathcal{D}_{-\nu}(z^\ast/\sqrt{2})}{\mathcal{D}_{-\nu}(z/\sqrt{2})}~.
\end{equation}
This ratio is, in general, non-thermal. This deviation from perfect thermality can be attributed to grey-body factors~\cite{Unruh1, Don}, which encode the fact that modes generated near the horizon do not propagate to the detector without modification. As they travel through the background geometry, portions of the wave packet are partially reflected, scattered, or otherwise distorted, leading to a frequency-dependent filtering of the detected spectrum.
Furthermore, for low-frequency outgoing modes ($f\simeq 0$), the ratio reduces to $\frac{\alpha_{ff^\prime}^G}{\beta_{ff^\prime}^G}\approx 1$, indicating an equal contribution from positive and negative-frequency components. As a consequence, the particle occupation number is given by
\begin{equation}
    n_f^{G}=\int _0^\infty df^\prime \,|\beta_{ff^\prime}|^2 = \frac{1}{2}~,
\end{equation}
which corresponds to the standard Fermi level.

\section{Qubit Dynamics and Decoherence}
\label{QubitAppendix}

\subsection{Exact Dynamics in Strong Coupling}

For the general coupling case, the reduced density matrix of the qubit $\rho_q(t)=\mathrm{Tr}_E \left[\,\ket{\Psi(t)}\bra{\Psi(t)}\,\right]$ is obtained by tracing over the spin-chain modes $E$. The matrix elements in the basis $\{\ket{g}, \ket{e}\}$ are given by:
\begin{align}
    P_{gg}(p,t) &= \frac{2g_1^2+g_2^2\left( \cos \left(\Omega_Rt\right)+1\right)}{2(g_1^2+g_2^2)}~, \nonumber \\
    P_{eg}(p,t) &= \frac{g_1g_2(\cos \left(\Omega_Rt\right)-1)}{2(g_1^2+g_2^2)}-i\frac{g_2\sin \left(\Omega_Rt\right)}{2\sqrt{g_1^2+g_2^2}}~,
\end{align}
where $P_{ee}(p,t)=1- P_{gg}(p,t)$ and $P_{ge}(p,t)=P_{eg}^\ast(p,t)$. The Rabi frequency is defined as $\Omega_R=2 \sqrt{g_1^2+g_2^2} \sin{2p}$. The total population is the product over all quasi-momenta $p$. The decoherence function~\cite{Liu}, defined as $D(t)=\prod_p|P_{ge}(p,t)|^2$, quantifies the magnitude of coherence present in the reduced density matrix. We emphasize that since the qubit starts in an energy eigenstate ($D(0)=0$ by construction), $D(t)$ measures the suppression of \textit{dynamically generated} coherence. {
As the qubit–chiral interaction is switched on, $D(t)$ initially increases, indicating the buildup of coherence. At later times, however, it saturates, reflecting the qubit completely entangled with the environmental degrees of freedom. This definition differs from commonly used decoherence measures, where $D(t)=1$ corresponds to a pure, fully coherent state and $D(t)=0$ signals complete loss of coherence.}

\begin{figure}[t]
     \centering
     \begin{subfigure}[b]{0.31\textwidth}
         \centering
         \includegraphics[width=\textwidth]{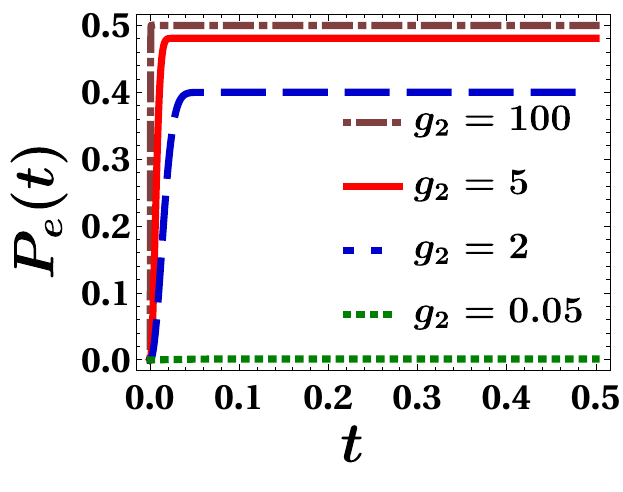}
         \caption{}
     \end{subfigure}
     \hfill
     \begin{subfigure}[b]{0.31\textwidth}
         \centering
         \includegraphics[width=\textwidth]{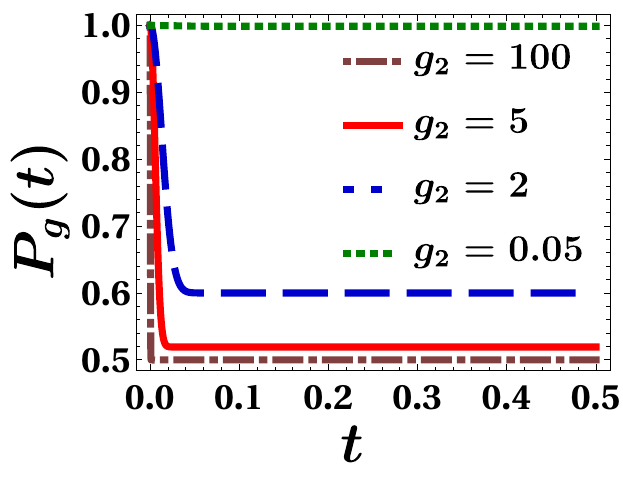}
         \caption{}
     \end{subfigure}
     \hfill
     \begin{subfigure}[b]{0.31\textwidth}
         \centering
         \includegraphics[width=\textwidth]{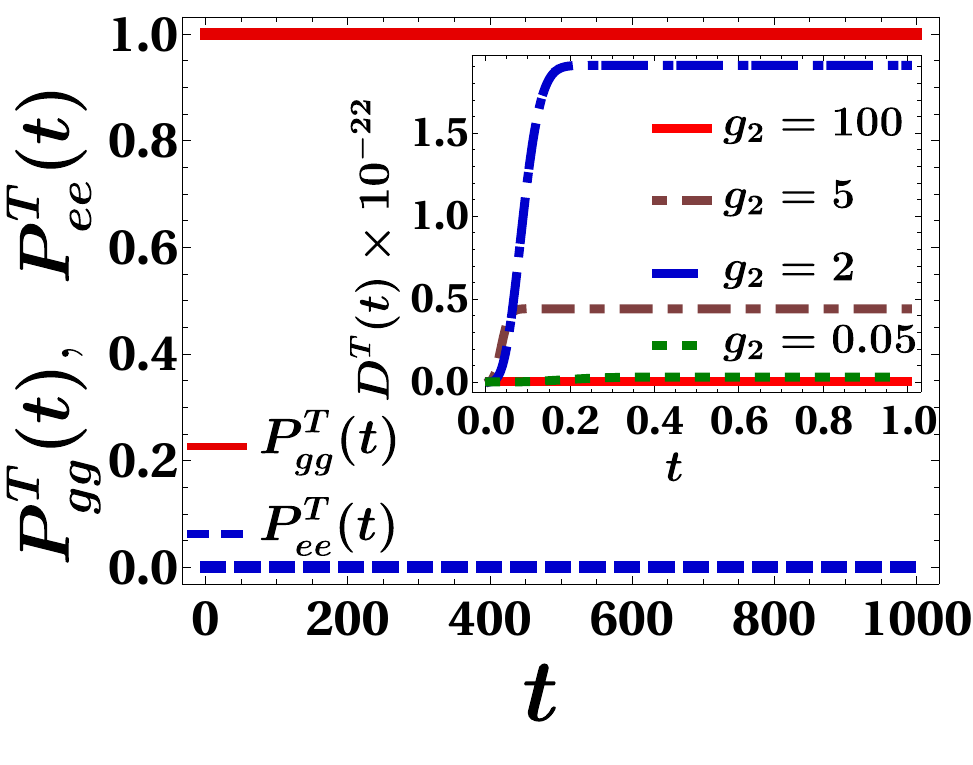}
         \caption{}
     \end{subfigure}
     \caption{Time evolution of the qubit populations for different values of the coupling strength $g_2$, with $g_1=1$ fixed. \textbf{(a)} Excited-state and \textbf{(b)} ground-state populations when the spin chain is initialized in its ground state. \textbf{(c)} Freezing of the qubit populations when the spin chain is prepared in a thermal state, signaling the suppression of coherent population transfer. The inset in (c) displays the corresponding decoherence factor induced by the thermal spin-chain environment.}
     \label{Populationfig}
\end{figure}

The left and middle panels of Fig.~\ref{Populationfig} show the time evolution of the qubit populations with the system initialized in the ground state ($P_{gg}(0)=1$). Although the combined system evolves unitarily, the spin chain acts as a finite many-body reservoir with internal excitations and correlations. Consequently, both phase and energy information of the qubit are rapidly redistributed into the spin-chain degrees of freedom via entanglement.

Due to this rapid decoherence and energy exchange, the qubit populations relax to steady-state values on a short timescale $\tau_{0}=\sqrt{N^{-1}\log{(\epsilon^{-1})}/(g_1^2+g_2^2)}$ (for accuracy $\epsilon=10^{-4}$). Beyond this timescale, the steady state can be characterized by an effective temperature:
\begin{equation}
    T_{\mathrm{ef}}=\frac{\left|\mathcal{U}-v\right|}{\log{\left(P_g^s/P_e^s\right)}}~,
\end{equation}
where $P_g^s$ and $P_e^s$ denote the steady-state ground and excited populations, respectively, and $|\mathcal{U}-v|$ is the energy exchange scale. 

We have explicitly verified that this effective temperature remains distinct from the Hawking temperature $T_H$, even when the spin chain is initialized in a thermal state $\rho_E \propto e^{-H_E/T}$. For such thermally disordered configurations, the qubit population dynamics are given by:
\begin{align}
     P_{gg}^T(t) &= \prod_p \left[1-\frac{g_2^2 f_p \left( 1-\cos \left(\Omega_Rt\right)\right)}{2(g_1^2+g_2^2)}\right]~,\quad f_p=\frac{1}{e^{\frac{E(p)}{T}}+1}~,\nonumber\\
     P_{eg}^T(t) &= \prod_p f_p\left[\frac{g_1g_2}{2(g_1^2+g_2^2)}\left( -1+\cos \left(\Omega_Rt\right)\right)-i\frac{g_2\sin \left(\Omega_Rt\right)}{2\sqrt{g_1^2+g_2^2}}\right]~.
\end{align}
In this thermal regime, the qubit populations remain effectively frozen ($P_{gg}^T(t)\approx 1$), as shown in Fig.~\ref{Populationfig}(c). The corresponding decoherence function $D^T(t)$ exhibits negligible decay (Fig.~\ref{Populationfig}(c) inset). This prevents meaningful temperature extraction, confirming that strong coupling is unsuitable for probing $T_H$.

\subsection{Weak Coupling Rates}

In the weak-coupling regime ($|g_2| \ll |\mathcal{U}|, |v|$), the qubit dynamics are governed by the Dirac spectrum of the chiral spin chain, which acts as an effective thermal reservoir. The reduced dynamics are described by the Born–Markov approximation, leading to rate equations determined entirely by the spectral properties of the environment. The relevant bath spectral density is defined as:
\begin{equation}
    J(\omega)=2\pi \left(g_1^2+g_2^2\right)\varrho(\omega), \quad \varrho(\omega)=\frac{1}{2\pi}\sum_{E(p)=\omega}\frac{1}{|\partial_pE(p)|}~,
\end{equation}
where $\omega$ denotes the energy exchange and $\varrho(\omega)$ is the density of states. For the dispersion relation considered here, this simplifies to $J(\omega)=(g_1^2+g_2^2)/|2(\mathcal{U}-v)|$. The transition rates are:
\begin{equation}
    \Gamma_\uparrow= J(\omega)f_D(\omega)~,\quad \Gamma_\downarrow= J(\omega)(1-f_D(\omega))~,
\end{equation}
where $f_D(\omega)=\left(e^{\omega/T}+1\right)^{-1}$. These rates yield the population dynamics of the excited state:
\begin{equation}
    \dot{\rho}_{ee}=- \Gamma_\downarrow \rho_{ee}+\Gamma_\uparrow (1-\rho_{ee}) \implies \rho_{ee}(t)=f_D(\omega)\left(1-e^{-J(\omega)t}\right)~.
\end{equation}
The ground-state population $\rho_{gg}(t)=1-\rho_{ee}(t)$ relaxes toward a stationary thermal state. This satisfies the detailed balance condition $\Gamma_\uparrow / \Gamma_\downarrow = e^{-\omega/T_H}$, validating the thermalization observed in Fig.~6 of the main text. 

At the horizon ($|\mathcal{U}|=|v|$), the bath spectral density diverges ($J(\omega)\rightarrow \infty$), signaling infinitely strong mode coupling. In this limit, the qubit states are fully governed by the Dirac thermal spectrum. Thus, as long as the analog black hole exists, the detector couples to the Dirac spectrum rather than the vacuum, allowing $T_H$ to be extracted as an emergent property of the environment.

\bibliography{lattice1}

\end{document}